# Intrinsic graphene field effect transistor on amorphous carbon films

S.S. Tinchev*

Institute of Electronics, Bulgarian Academy of Sciences,

Sofia 1784, Bulgaria

Abstract:

Fabrication of graphene field effect transistor is described which uses an intrinsic graphene on the surface of as deposited hydrogenated amorphous carbon films. Ambipolar characteristic has been demonstrated typical for graphene devices, which changes to unipolar characteristic if the surface graphene was etched in oxygen plasma. Because amorphous carbon films can be growth easily, with unlimited dimensions and no transfer of graphene is necessary, this can open new perspective for graphene electronics.

*e-mail: stinchev@ie.bas.bg

Recently [1] we have measured surface resistivity of hydrogenated amorphous carbon films and observed that the surface resistivity shows a sharp ambipolar peak at +0.3 V of the gate voltage and sheet resistance at the peak maximum of 7.5 k$\Omega$/sq. This value is the same order of magnitude as the sheet resistance of a defect free graphene monolayer [2]. Therefore a conclusion was made that an intrinsic graphene exist on the surface of hydrogenated amorphous carbon films and probably on the surface of all other types of amorphous carbon films. In this letter we report on field effect transistor fabricated from intrinsic graphene, which exist on the surface of as deposited amorphous carbon films.

The films used in our experiments were 70 nm amorphous hydrogenated (a-C:H) carbon films deposited by PE CVD (plasma enhanced CVD) from benzene vapor diluted with argon. They are fabricated on the top of 300 nm thermal $SiO_2$ on silicon. The $SiO_2$ serves as an insulating layer, so a back-gate voltage can be applied to vary carrier concentration. These films can be produced with very different resistivity from soft graphitic-like low resistance films to hard, high resistance films by varying only the bias voltage in a DC PE CVD system. As we needed high resistance samples the films used in this work were fabricated at 1kV bias voltage.

Fig. 1 shows the field effect transistor structure. The conducting channel is formed at the top of the surface of amorphous carbon film between two silver paint contacts serving as source and drain of the transistor. Silver paint contacts were used instead of evaporated or sputtered metal contacts in order to prevent possible modification of the film surface during contact fabrication. The same reason was to use fabrication process without patterning the film. The distance between source and drain contacts was about 3 mm.

The back-gate voltage from a low frequency signal generator or from a digital –analog converter was applied to the back side of the silicon substrate through a silver paint contact. In this measurement the silicon substrate is acting as a gate and 300 nm $SiO_2$ is a gate insulator. Current – voltage characteristics of the fabricated transistors were measured swiping a DC current from a digital – analog converter. The source – drain voltage was recorded by an analog-digital converter with 13 bit resolution.

Current –voltage characteristic of intrinsic graphene field effect transistor is shown in Fig. 2. It is found that this device exhibit a nearly linear I-V dependence and lack of current saturation. Clear modulation of the source – drain resistance by the gate voltage is visible. In these measurements the gate voltage was ramped from -1 V to 1 V with frequency 1 Hz. The measured characteristic is not full symmetrical probably due to non-identical source and drain contacts. Indeed in other samples we observed symmetrical current-voltage characteristics with almost identical positive and negative branches.

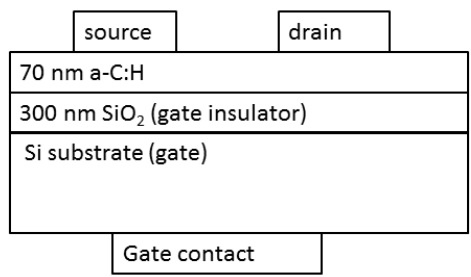

Fig. 1. Graphene field effect transistor structure

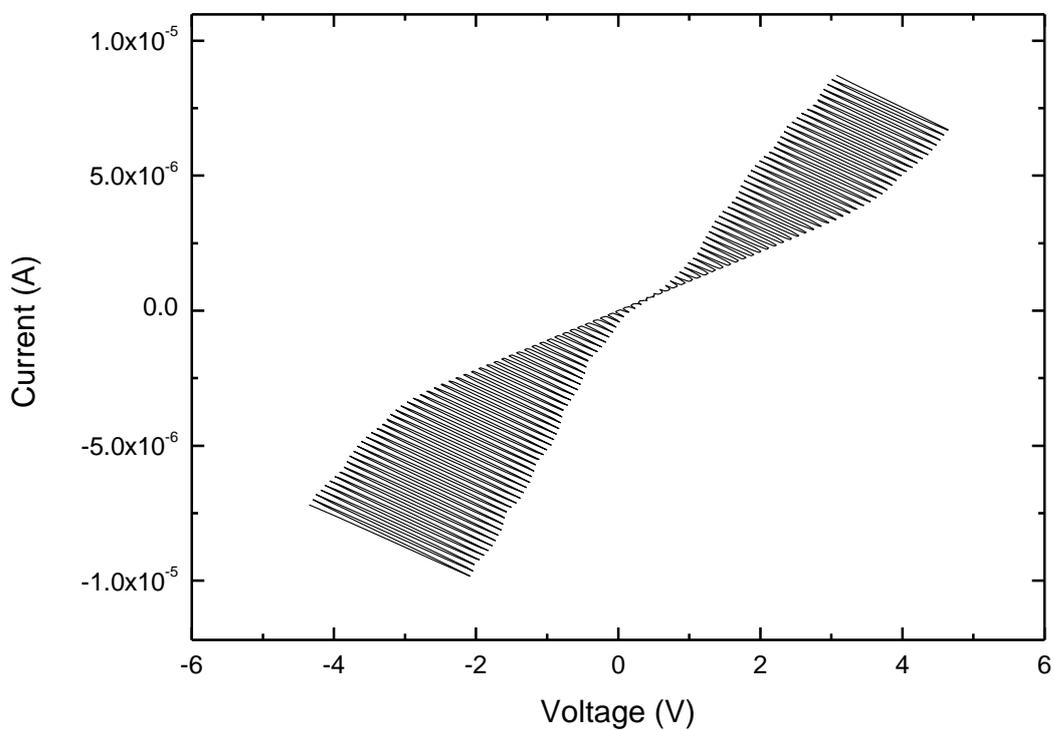

Fig. 2. Current – voltage characteristic of intrinsic graphene field effect transistor.

The measured transfer characteristic - Fig. 3 (drain current versus gate voltage) of the transistor is ambipolar, although unipolar characteristics are often reported in graphene transistors [3-5] and are explained usually by p-dopants, such as the oxygen and water adsorbed on graphene surfaces. For gate voltages between -0.8 V and +2.5 V the current is constant, about 0.3 µA probably flowing though the bulk of the DLC film. For negative gate voltages greater then -0.8 V and for positive voltages greater then +2.5 V the current increases, which is a clear indication of an ambipolar behavior. The branch of this characteristic for positive gate voltages is steeper because of the greater mobility of electrons carrying the current in this branch. For dc drain-source voltage of 0.191 V the transconductance of the transistor is 1.6 µS for holes and 2.6 µS for electrons. Because the channel dimensions are not well defined this value is difficult to compare with other graphene transistors.

In the next experiment the graphene layer on the surface of amorphous carbon was removed by plasma etching. The etching was made in pulsed oxygen plasma [1] for processing time of 10 s at voltage amplitude 740 V, pulse frequency of 66 kHz and pulse time of 10 µs. During the plasma etching the pressure of the chamber was $2.6 \times 10^{-1}$ Torr. Transfer characteristic of a field effect transistor fabricated from this film (shown in Fig. 4) is unipolar and typical for p-channel mode of operation of a field effect transistor. This can be expected from the p-type nature of the hydrogenated amorphous carbon films.

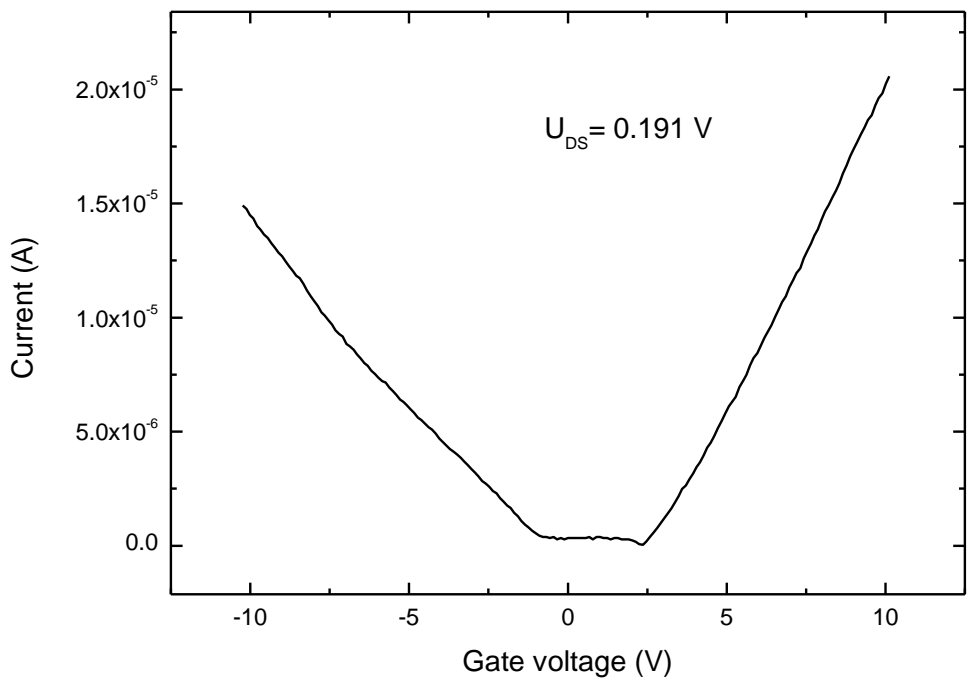

Fig. 3. Transfer characteristic of intrinsic graphene field effect transistor with an applied drain voltage of 0.191 V.

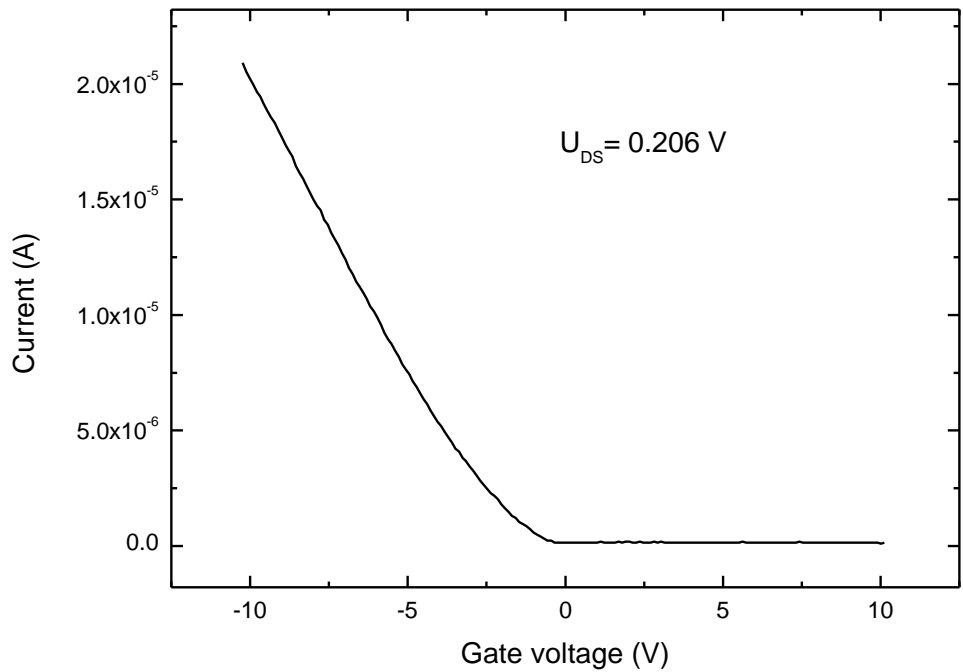

Fig. 4. Transfer characteristic of field effect transistor fabricated after removing graphene surface layer by oxygen plasma. The applied drain voltage is 0.206 V.

Here one should remember that by using transmission electron microscopy (TEM) and spatially resolved electron energy loss spectroscopy (EELS) $sp^2$-bonded layer about 1 nm thick was already observed at the surface of the tetrahedral amorphous carbon films [6, 7]. The existence of this $sp^2$-rich layer on the surface of tetrahedral (ta-C) amorphous carbon was regarded as problem for fabrication of TFT (thin film) transistors from these materials. One believes that this layer could prevent transistor action and has to be removed. In [8] the graphitic layer on the top surface of the ta-C was etched away in oxygen plasma and ta-C thin film transistor was fabricated.

In conclusion, graphene field effect transistor has been fabricated using intrinsic graphene, which exists on the surface of as deposited hydrogenated amorphous carbon films. Gate control of the drain current has been demonstrated in ambipolar transfer characteristics. For dc drain-source voltage of 0.191 V the transconductance of the transistor is 1.6 µS for holes and 2.6 µS for electrons. Removing the graphene surface layer in oxygen plasma changes the ambipolar behavior to unipolar. Because amorphous carbon films can be growth easily, with unlimited dimensions and no transfer of grapheneis is necessary, it can open new perspective for graphene electronics.